\documentstyle[epsf]{elsart}
\begin{document}

\begin{frontmatter}

\title{
Effects of the ground state correlations on the structure
of vibrational states
}
\author{D. Karadjov$^a$, V.V. Voronov$^b$, F. Catara$^c$,
M. Grinberg$^a$, A.P. Severyukhin$^b$}
\address{
$^a$Institute for Nuclear Research and Nuclear Energy, 1784 Sofia,
Bulgaria\\
$^b$ - Bogoliubov Laboratory of Theoretical Physics, JINR, 141980 Dubna,
Russia\\
$^c$ - Dipartimento di Fisica dell' Universita' and INFN, Catania, Italy
}
\maketitle

\begin{abstract}
A method to treat the ground state
correlations beyond the RPA is presented. A
set of nonlinear equations taking into
account effects of the ground state
correlations on the pairing and
phonon-phonon coupling is derived. The
influence of such correlations on properties
of the vibrational states in spherical
nuclei is studied.
\end{abstract}

\begin{keyword}
extended random phase approximation;
vibrational states; charge transition
density; Quasiparticle Phonon Model
calculation. \\ {\it PACS:} 21.60.-n;
21.60.Jz; 23.20.-g \\
\end{keyword}

\end{frontmatter}

\section{Introduction.}

The Random Phase Approximation (RPA) is the basic method to describe the
nuclear vibrational motion. It is well known that the RPA violates the Pauli
principle and many attempts have been performed to improve it (\cite
{H,IUY65,R1,R2,Pr68,JR,NR,JM88,LW90,KWD91,KVC93,KVC94,CDS94,SC95,DS96}).
There is a rather complete list of references on that subject in paper \cite
{KWD91}. Renormalized RPA equations that include corrections for the ground
state correlations (GSC), have been applied not only to study properties of
the low-lying isoscalar vibrations in spherical nuclei \cite{KVC93,KVC94},
but also to investigate the charge-exchange modes in nuclei \cite{TS97,HHC97}
and the giant resonances in metal clusters \cite{CPSG96}. To describe
excited states all above mentioned works use the one phonon states that can
be created acting by the relevant collective operator on the ground state of
the system, which is treated as the vacuum state. From another point of view
it is well known that due to the anharmonicity of vibrations there is a
coupling between one-phonon and more complex states \cite{BM75,Sol92}.
Taking into account such a coupling it is possible to describe
particularities of the low-lying states and damping of the giant resonances
\cite{Sol92}. Up to now such coupling was considered for the RPA phonons
only \cite{Sol92}.

In the present paper we use phonons of the extended RPA (ERPA) \cite{KVC94}
as a basis on which the quasi-particle phonon model (QPM) \cite{Sol92}
equations are generalized so as to account for the GSC in the description of
nuclear vibrational states constructed by one- and two-phonon
configurations. Besides the GSC, we take into account the Pauli principle
corrections arising in the two-phonon terms due to the fermion structure of
the phonon operators. As an example we study the effect of the GSC on the
energies, transition probabilities and transition densities of the low-lying
vibrational states in $^{68}$Zn and compare present results with the results
within other approaches.

{\bf 2. Basic formulae}\vspace{1mm}

We employ the QPM - Hamiltonian including an average nuclear field described
as the Woods-Saxon potential, pairing interactions, the isoscalar and
isovector particle--hole (p--h) residual forces in separable form with the
Bohr--Mottelson radial dependence \cite{BM75}:
\begin{eqnarray}
H=\sum\limits_\tau \{\sum\limits_{jm}(E_j-\lambda_\tau )a_{jm}^{\dagger
}a_{jm}-\frac 14G_\tau ^{(0)}:(P_0^{\dagger }P_0)^\tau :-  \nonumber \\
\frac 12\sum\limits_{\lambda \mu \sigma=\pm1}(\kappa_{0}^{(\lambda )}+
\sigma\kappa_{1}^{(\lambda)}) :(M_{\lambda \mu }^{\dagger}(\tau)M_{\lambda
\mu }(\sigma \tau)):\}  \label{eq1}
\end{eqnarray}
We sum over the proton($p$) and neutron($n$) indexes and the notation $%
\{\tau =(n,p)\}$ is used and a change $\tau \leftrightarrow -\tau $ means a
change $p\leftrightarrow n$. The single-particle states are specified by the
quantum numbers $(jm)$; $E_j$ are the single-particle energies; $\lambda
_\tau $ is the chemical potential; $G_\tau ^{(0)}$ and $\kappa ^{(\lambda )}$
are the strengths in the p--p and in the p--h channel, respectively. The
pair creation and the multipole operators entering the normal products in
(1) are defined as follows:
\begin{equation}
P_0^{+}\,=\,\sum_{jm}(-1)^{j-m}a_{jm}^{+}a_{j-m}^{+}  \label{eq2}
\end{equation}
\begin{equation}
M_{\lambda \mu }^{+}\left( \tau \right) \,=\,\frac 1{\sqrt{2\lambda +1}%
}\sum_{jj^{^{\prime }}mm^{^{\prime }}}^{\left( \tau \right)
}(-1)^{j+m}\langle jmj^{^{\prime }}-m^{^{\prime }}\mid \lambda \mu \rangle
f_{j^{^{\prime }}j}^{(\lambda )}a_{jm}^{+}a_{j^{^{\prime }}m^{^{\prime }}}
\label{eq3}
\end{equation}
where $f_{jj^{^{\prime }}}^{(\lambda )}$ are the single particle radial
matrix elements of residual forces.

In what follows we work in quasiparticle (qp) representation, defined by the
canonical Bogoliubov transformation:
\begin{equation}
a_{jm}^{+}\,=\,u_j\alpha _{jm}^{+}\,+\,(-1)^{j-m}v_j\alpha _{j-m}
\label{eq4}
\end{equation}
The Hamiltonian can be represented in terms of bifermion quasiparticle
operators (and their conjugate ones):
\begin{equation}
B(jj^{^{\prime }};\lambda \mu )\,=\,\sum_{mm^{^{\prime }}}(-1)^{j^{^{\prime
}}+m{^{\prime }}}\langle jmj^{^{\prime }}m^{^{\prime }}\mid \lambda \mu
\rangle \alpha _{jm}^{+}\alpha _{j^{^{\prime }}-m^{^{\prime }}}  \label{eq5}
\end{equation}
\begin{equation}
A^{+}(jj^{^{\prime }};\lambda \mu )\,=\,\sum_{mm^{^{\prime }}}\langle
jmj^{^{\prime }}m^{^{\prime }}\mid \lambda \mu \rangle \alpha
_{jm}^{+}\alpha _{j^{^{\prime }}m^{^{\prime }}}^{+}  \label{eq6}
\end{equation}

The phonon creation operators are defined in the 2-qp space in a standard
fashion:
\begin{equation}
Q_{\lambda \mu ,i}^{+}\,=\,\frac 12\sum_{jj^{^{\prime }}}\{\psi
_{jj^{^{\prime }}}^{\lambda i}\,A^{+}(jj^{^{\prime }};\lambda \mu
)-(-1)^{\lambda -\mu }\varphi _{jj^{^{\prime }}}^{\lambda i}\,A(jj^{^{\prime
}};\lambda -\mu )\}  \label{eq7}
\end{equation}
where the index $\lambda =0,1,2,3,...$ denotes multipolarity and $\mu $ is
its z-projection in the laboratory system.The following relation can be
proved using the exact commutators of the fermion operators:
\begin{equation}
\langle 0\mid \,[Q_{\lambda \mu ,i},Q_{\lambda ^{^{\prime }}\mu ^{^{\prime
}},i^{^{\prime }}}^{+}]\,\mid 0\rangle \,=\,\frac 12\delta _{\lambda \lambda
^{^{\prime }}}\delta _{\mu \mu ^{^{\prime }}}\,\sum_{jj^{^{\prime
}}}\,(1\,-q_{jj^{^{\prime }}})[\psi _{jj^{^{\prime }}}^{\lambda i}\,\psi
_{jj^{^{\prime }}}^{\lambda i^{^{\prime }}}\,-\,\varphi _{jj^{^{\prime
}}}^{\lambda i}\varphi _{jj^{^{\prime }}}^{\lambda i^{^{\prime }}}]
\label{eq8}
\end{equation}
where $\mid 0\rangle $ is the phonon vacuum, $q_{jj^{^{\prime
}}}=q_j+q_{j^{^{\prime }}}$ and $q_j$ is the quasiparticle distribution in
the ground state: $q_j\equiv (2j+1)^{-\frac 12}\langle 0\mid
\,B(jj;00)\,\mid 0\rangle$.

The phonon and the pairing characteristics are determined by the following
non--linear system of equations:
\begin{equation}
\sum_\tau \left[ \left( k_0^{\left( \lambda \right) }+k_1^{\left( \lambda
\right) }\right) X_\tau ^{\lambda i}-2k_0^{\left( \lambda \right)
}k_1^{\left( \lambda \right) }X_\tau ^{\lambda i}X_{-\tau }^{\lambda
i}\right] =1  \label{eq9}
\end{equation}

\begin{equation}
X_\tau ^{\lambda i}=\,\sum_{jj^{^{\prime }}}^{(\tau )}\,\frac{
(f_{jj^{^{\prime }}}^{(\lambda )}u_{jj^{^{\prime }}}^{\left( +\right)
})^2\,\varepsilon _{jj^{^{\prime }}}\,(1\,-\,q_{jj^{^{\prime }}})}{%
\varepsilon _{jj^{^{\prime }}}^2\,-\,\omega _{\lambda i}^2}  \label{eq10}
\end{equation}

\begin{equation}
\sum_{jj^{^{\prime }}}\,(1\,-q_{jj^{^{\prime }}})\,[(\psi _{jj^{^{\prime
}}}^{\lambda i})^2\,-\,(\varphi _{jj^{^{\prime }}}^{\lambda
i})^2]\,-\,2\,=\,0  \label{eq11}
\end{equation}

\begin{equation}
\frac{G_\tau ^{\left( 0\right) }}2\sum\limits_j^{(\tau )}\frac{%
(j\,+\,1/2)(1\,-\,2q_j)}{\sqrt{\Delta _\tau ^2\,+\,(E_j\,-\,\lambda _\tau
)^2 }}=1  \label{eq12}
\end{equation}

\begin{equation}
\sum\limits_j^{(\tau )}(j\,+\,1/2)\left[ 1-\frac{%
(E_j\,-\,\lambda _\tau )\,(1\,-\,2q_j)}{\sqrt{\Delta _\tau
^2\,+\,(E_j\,-\,\lambda _\tau )^2}}\right]=\,N^{(\tau )} \label{eq13}
\end{equation}

\begin{equation}
q_j\,=\,\frac 12\sum_{\lambda i,j^{^{\prime }}}\,\frac{2\lambda \,+\,1}{%
2j\,+\,1}\,(1\,-q_{jj^{^{\prime }}})\,(\varphi _{jj^{^{\prime }}}^{\lambda
i})^2  \label{eq14}
\end{equation}

The formulae for the quasiparticle energies $\varepsilon _j=\sqrt{\Delta
_\tau ^2\,+\,(E_j\,-\,\lambda _\tau )^2}$ and for the coefficients $%
u_j,\,v_j $ remain the same as in the usual BCS theory; the new values for $%
\Delta _\tau \equiv \frac 12G_\tau ^{(0)}\sum\limits_j^{(\tau
)}(1-2q_j)(2j+1)u_j\,v_j$ and $\lambda _\tau $, come from the eqs. (12) and
(13); $\varepsilon _{jj^{^{\prime }}}=\varepsilon _j+\varepsilon
_{j^{^{\prime }}},$ $u_{jj^{^{\prime }}}^{(+)}=u_jv_{j^{^{\prime
}}}+v_ju_{j^{^{\prime }}}$. The pairing vibrations ($\lambda=0$) have been
considered in \cite{KVC96}.

The system of nonlinear equations (9)-(14) includes effects of the isoscalar
and isovector forces and it is a generalization of equations derived in \cite
{H,KVC93,KVC94,VK95}. This system treats the GSC self-consistently and
describes the coupling between different vibrations, different phonon roots
of a certain multipolarity and the pairing. The present scheme is called
Extended RPA (ERPA). The factors $(1\,-\,q_{jj^{^{\prime }}}),$
distinguishing the new equations from the conventional BCS and RPA ones,
take into account the blocking effect due to the Pauli principle. If we put $%
q_{jj^{\prime}}=0$ in the r.h.s. of eq. (14), we get the expression for the
quasiparticle distribution in the ground state in the RPA case \cite{R1,LW90}.

The GSC affect not only the RPA, but they also should change the
quasiparticle-phonon coupling. To take into account such effects we follow
the basic ideas of the QPM. Hereafter we derive the generalized QPM
equations which take into account the GSC beyond the RPA. As it was shown in
our previous paper \cite{KVC96} the pairing vibrations give a negligible
contribution to $q_j$. On the other hand the two-phonon configurations
including the pairing vibration phonons have an energy essentially higher
than the configurations constructed from usual vibration phonons. That is
why we do not take into account the coupling with the pairing vibrations in
what follows. Using the completeness and orthogonality conditions for the
phonon operators one can express bifermion operators $A^{+}$ and $A$ by
phonons:

\begin{eqnarray}
A^{+}(jj{^{\prime }};\lambda \mu )\,+\,(-)^{\lambda -\mu }\,A(jj^{^{\prime
}};\lambda -\mu )\,=  \nonumber \\
\,(1\,-\,q_{jj^{^{\prime }}})\sum_i(\psi _{jj^{^{\prime }}}^{\lambda
i}\,+\,\phi _{jj^{^{\prime }}}^{\lambda i})(Q_{\lambda \mu
i}^{+}\,+\,(-)^{\lambda -\mu }\,Q_{\lambda -\mu i})  \label{eq15}
\end{eqnarray}

The initial Hamiltonian (1) can be rewritten in terms of quasiparticle and
phonon operators in following form:

\begin{equation}
H\,=\,h_{0}\,+\,h_{pp}\,+\,h_{QQ}\,+\,h_{QB}  \label{eq16}
\end{equation}

\begin{equation}
h_0\,+\,h_{pp}\,=\,\sum_{jm}\,\varepsilon _j\,\alpha _{jm}^{+}\,\alpha _{jm}
\label{eq17}
\end{equation}

\begin{equation}
h_{QQ}=-\frac 18\sum_{\lambda \mu ii^{^{\prime }}\tau }\frac{X_\tau
^{\lambda i}+X_\tau ^{\lambda i^{^{\prime }}}}{\sqrt{{\cal Y}_\tau ^{\lambda
i}{\cal Y}_\tau ^{\lambda i^{^{\prime }}}}}(Q_{\lambda \mu
i}^{+}+(-)^{\lambda -\mu }Q_{\lambda -\mu i})(Q_{\lambda -\mu i^{^{\prime
}}}^{+}+(-)^{\lambda +\mu }Q_{\lambda \mu i^{^{\prime }}})  \label{eq18}
\end{equation}

\begin{equation}
h_{QB}=-\frac 1{2\sqrt{2}}\sum_{\lambda \mu ijj^{^{\prime }}\tau }\frac{%
v_{jj^{^{\prime }}}^{(-)}f_{jj^{^{\prime }}}^{(\lambda )}}{\sqrt{{\cal Y}
_\tau ^{\lambda i}}}((-)^{\lambda -\mu }Q_{\lambda \mu i}^{+}+Q_{\lambda
-\mu i})B_\tau (jj^{^{\prime }};\lambda -\mu )+h.c.  \label{eq19}
\end{equation}

where $v_{jj^{^{\prime }}}^{(-)}\,=\,u_ju_{j^{^{\prime
}}}\,-\,v_jv_{j^{\prime }}$ and

\begin{equation}
{\cal Y}_\tau ^{\lambda i}=Y_\tau ^{\lambda i}+Y_{-\tau }^{\lambda i}\left\{
\frac{1-\left( k_0^{\left( \lambda \right) }+k_1^{\left( \lambda \right)
}\right) X_\tau ^{\lambda i}}{\left( k_0^{\left( \lambda \right)
}-k_1^{\left( \lambda \right) }\right) X_{-\tau }^{\lambda i}}\right\} ^2
\label{eq20}
\end{equation}
\begin{equation}
Y_\tau ^{\lambda i}=\,\sum_{jj^{^{\prime }}}^{\left( \tau \right) }\,\frac{
(f_{jj^{^{\prime }}}^{(\lambda )}u_{jj^{^{\prime }}}^{\left( +\right)
})^2\,\varepsilon _{jj^{^{\prime }}}\,\omega _{\lambda
i}(1\,-\,q_{jj^{^{\prime }}})}{[\varepsilon _{jj^{^{\prime }}}^2\,-\,\omega
_{\lambda i}^2]^2}  \label{eq21}
\end{equation}
One can prove that the solutions of the system of equations (9)--(14) obey
the following equality:

\begin{equation}
\langle \,Q_{\lambda \mu i}\,\mid \,H\,\mid \,Q_{\lambda \mu i}^{+}\rangle
\,=\,\omega _{\lambda i}  \label{eq22}
\end{equation}

The term $h_{QB}$ is responsible for the mixing of the configurations and,
therefore, for the description of many characteristics of the excited states
of even--even nuclei. In the simplest case the wave functions of those
states could be written down as:

\begin{equation}
\Psi _\nu (\lambda \mu )=\{\sum_iR_i(\lambda \nu )Q_{\lambda \mu
i}^{+}+\sum_{\lambda _1i_1\lambda _2i_2}P_{\lambda _1i_1}^{\lambda
_2i_2}(\lambda \nu )\left[ Q_{\lambda _1\mu _1i_1}^{+}Q_{\lambda _2\mu
_2i_2}^{+}\right] _{\lambda \mu }\}|0\rangle  \label{eq23}
\end{equation}

with the normalization condition:

\begin{eqnarray}
\langle \Psi _\nu (JM) \mid \Psi _\nu (JM)\rangle = \sum\limits_iR_i^2(J\nu
)+  \nonumber \\
2\sum_{\lambda _1i_1 \lambda _2i_2} (P_{\lambda _2i_2}^{\lambda _1i_1}(J\nu
))^2(1+K^J(\lambda _1i_1,\lambda _2i_2))=1  \label{eq24}
\end{eqnarray}

where

\[
K^J(\lambda _1i_1,\lambda _2i_2)\equiv K^J(\lambda _1i_1,\lambda _2i_2\mid
\lambda _2i_2,\lambda _1i_1)
\]

and

\begin{eqnarray}
K^J(\lambda _2i_2,\lambda i^{^{\prime }} \mid \lambda i,\lambda
_2i_2)=(2\lambda +1)(2\lambda _2+1)\frac 1{1+\delta _{i,i^{^{\prime
}}}\delta _{\lambda i,\lambda _2i_2}}\times  \nonumber \\
\sum\limits_{j_1j_2j_3j_4}(1-\frac
12q_{j_1j_2j_3j_4})(-1)^{j_2+j_4+J}\left\{
\begin{array}{ccc}
j_1 & j_2 & \lambda _2 \\
j_4 & j_3 & \lambda \\
\lambda & \lambda _2 & J
\end{array}
\right\} \times  \nonumber \\
(\psi _{j_1j_4}^{\lambda i^{^{\prime }}}\psi _{j_3j_4}^{\lambda i}\psi
_{j_3j_2}^{\lambda _2i_2}\psi _{j_1j_2}^{\lambda _2i_2}-\varphi
_{j_3j_2}^{\lambda _2i_2}\varphi _{j_1j_2}^{\lambda _2i_2}\varphi
_{j_3j_4}^{\lambda i^{^{\prime }}}\varphi _{j_1j_4}^{\lambda i})  \label{25}
\end{eqnarray}

\[
q_{j_1j_2j_3j_4} \equiv q_{j_1}+q_{j_2}+q_{j_3}+q_{ j_4}
\]

The mean value of $H$ is

\begin{eqnarray}
\langle \Psi _\nu (JM) &\mid &H\mid \Psi _\nu (JM)\rangle
=\sum\limits_iR_i^2(J\nu )\omega _{Ji}+2\sum_{\lambda _1i_1\lambda
_2i_2}(P_{\lambda _2i_2}^{\lambda _1i_1}(J\nu ))^2\times  \nonumber \\
&&(\omega _{\lambda _1i_1}+\omega _{\lambda _2i_2}+\Delta \omega ^J(\lambda
_1i_1,\lambda _2i_2))(1+K^J(\lambda _1i_1,\lambda _2i_2))+  \nonumber \\
&&2\sum_{\lambda _1i_1,i\lambda _2i_2}R_i(J\nu )P_{\lambda _2i_2}^{\lambda
_1i_1}(J\nu )U_{\lambda _2i_2}^{\lambda _1i_1}(J\nu )(1+K^J(\lambda
_1i_1,\lambda _2i_2))  \label{eq26}
\end{eqnarray}

where the matrix element coupling one- and two-phonon configurations is:

\begin{equation}
\langle Q_{J\nu }|h_{QB}|\left[ Q_{\lambda _1i_1}^{+}Q_{\lambda
_2i_2}^{+}\right] _J\rangle =U_{\lambda _2i_2}^{\lambda _1i_1}(J\nu
)(1+K^J(\lambda _1i_1,\lambda _2i_2))  \label{27}
\end{equation}

\[
U_{\lambda _2i_2}^{\lambda _1i_1}(\lambda i)=(-1)^{\lambda _1+\lambda
_2+\lambda }U_{\lambda _1i_1}^{\lambda _2i_2}(\lambda i)
\]
\[
U_{\lambda _2i_2}^{\lambda _1i_1}(\lambda i)\equiv \sum_\tau U_{\lambda
_2i_2}^{\lambda _1i_1}(\lambda i,\tau )
\]

\begin{eqnarray}
U_{\lambda _2i_2}^{\lambda _1i_1}(\lambda i,\tau ) &=&\frac 1{\sqrt{2}}\sqrt{%
(2\lambda _1+1)(2\lambda _2+1)}\sum_{j_1j_2j_3}^{\left( \tau \right)
}(1-q_{j_2j_3})\times  \nonumber \\
&&\,(\frac{f_{j_1j_2}^\lambda v_{j_1j_2}^{(-)}}{\sqrt{{\cal Y}_\tau
^{\lambda i}}}\left\{
\begin{array}{ccc}
\lambda _1 & \lambda _2 & \lambda \\
j_2 & j_1 & j_3
\end{array}
\right\} (\psi _{j_2j_3}^{\lambda _2i_2}\phi _{j_3j_1}^{\lambda _1i_1}+\psi
_{j_3j_1}^{\lambda _1i_1}\phi _{j_2j_3}^{\lambda _2i_2})+  \nonumber \\
&&\frac{f_{j_1j_2}^{\lambda _1}v_{j_1j_2}^{(-)}}{\sqrt{{\cal Y}_\tau
^{\lambda _1i_1}}}\left\{
\begin{array}{ccc}
\lambda _1 & \lambda _2 & \lambda \\
j_3 & j_2 & j_1
\end{array}
\right\} (\phi _{j_3j_1}^{\lambda _2i_2}\phi _{j_2j_3}^{\lambda i}+\psi
_{j_2j_3}^{\lambda i}\psi _{j_3j_1}^{\lambda _2i_2})+  \nonumber \\
&&\frac{f_{j_1j_2}^{\lambda _2}v_{j_1j_2}^{(-)}}{\sqrt{{\cal Y}_\tau
^{\lambda _2i_2}}}\left\{
\begin{array}{ccc}
\lambda _1 & \lambda _2 & \lambda \\
j_1 & j_3 & j_2
\end{array}
\right\} (\phi _{j_2j_3}^{\lambda _1i_1}\phi _{j_3j_1}^{\lambda i}+\psi
_{j_3j_1}^{\lambda i}\psi _{j_2j_3}^{\lambda _1i_1}))  \label{eq28}
\end{eqnarray}
and
\begin{eqnarray}
\Delta \omega ^J(\lambda _1i_1,\lambda _2i_2) &=&-\frac 14\sum_{i\tau }[
\frac{X_\tau ^{\lambda _1i_1}+\,X_\tau ^{\lambda _1i}}{\sqrt{{\cal Y}_\tau
^{\lambda _1i_1}{\cal Y}_\tau ^{\lambda _1i}}}K^J(\lambda _2i_2,\lambda
_1i\mid \lambda _1i_1,\lambda _2i_2)+  \nonumber \\
\frac{X_\tau ^{\lambda _2i_2}+\,X_\tau ^{\lambda _2i}}{\sqrt{{\cal Y}_\tau
^{\lambda _2i_2}{\cal Y}_\tau ^{\lambda _2i}}}K^J(\lambda _2i,\lambda _1i_1
&\mid &\lambda _1i_1,\lambda _2i_2)]  \label{eq29}
\end{eqnarray}

Calculating the mean value of $H$ we used so called quasidiagonal
approximation for the quantities $K^J(\lambda _1i_1,\lambda _2i_2\mid
\lambda _3i_3,\lambda _4i_4)$ because the diagonal terms dominate over
nondiagonal ones (see \cite{Sol92,VS83}).

Using the variational principle in the form:

\begin{equation}
\delta {\ \langle \,\Psi _\nu (\lambda \mu )\,|H|\,\Psi _\nu (\lambda \mu
)\rangle \,-\,E}_\nu {(\langle \,\Psi _\nu (\lambda \mu )\,|\,\Psi _\nu
(\lambda \mu )\rangle \,-\,1)\,}\,=\,0  \label{eq30}
\end{equation}
one obtains the following system of equations:

\begin{equation}
(\omega _{Ji}-E_\nu )R_i(J\nu )+\sum_{\lambda _1i_1 \lambda_2i_2} P_{\lambda
_2i_2}^{\lambda _1i_1}(J\nu )U_{\lambda _2i_2}^{\lambda _1i_1}(J\nu
)(1+K^J(\lambda _1i_1,\lambda _2i_2))=0  \label{eq31}
\end{equation}

\begin{equation}
2(\omega _{\lambda _1i_1}+\omega _{\lambda _2i_2}+\Delta \omega ^J(\lambda
_1i_1,\lambda _2i_2)-E_\nu )P_{\lambda _2i_2}^{\lambda _1i_1}(J\nu
)+\sum\limits_iR_i(J\nu )U_{\lambda _2i_2}^{\lambda _1i_1}(Ji)=0  \label{32}
\end{equation}

The energies of the states (23) are solutions of

\begin{equation}
F(E_\nu )\equiv det\left| (\omega _{\lambda i}-E_\nu )\delta _{ii^{\prime
}}-\frac 12\sum_{\lambda _1i_1,\lambda _2i_2}\frac{U_{\lambda
_2i_2}^{\lambda _1i_1}(\lambda i)U_{\lambda _2i_2}^{\lambda _1i_1}(\lambda
i^{\prime })\left( 1+K^J(\lambda _1i_1,\lambda _2i_2)\right) }{\omega
_{\lambda _1i_1}+\omega _{\lambda _2i_2}+\Delta \omega ^J(\lambda
_1i_1,\lambda _2i_2)-E_\nu }\right| =0  \label{33}
\end{equation}

The rank of the determinant (33) is determined by the number of the
one-phonon configurations included in the first term of the wave function
(23).

The above derived equations have the same form as the basic QPM-equations
\cite{Sol92,VS83} and we call them the extended QPM with the Pauli principle
corrections (EQPMPP) in what follows. The GSC affect phonon energies $\omega
_{\lambda i}$, normalization constants ${\cal Y}_{\tau}^{\lambda i}$ and
renormalize the matrix elements of the quasiparticle-phonon interaction. If
we put $K^J(\lambda _1i_1,\lambda _2i_2)=0$ we get the equations already
derived in\cite{VK95} where all the fourth-order terms in phonon amplitudes $%
\psi _{jj^{^{\prime }}}^{\lambda i}$ and $\phi _{jj^{^{\prime }}}^{\lambda
i} $ were neglected (we call this approach the extended QPM (EQPM)). In the
limit $q_{jj^{^{\prime }}}\rightarrow 0$ one reproduces precisely all the
expressions of the QPM with taking into account the Pauli principle
corrections \cite{Sol92,VS83}(QPMPP). In the case when $q_{jj^{^{\prime
}}}=0 $ and $K^J(\lambda _1i_1,\lambda _2i_2)=0$ we have equations
describing coupling of one- and two- RPA phonons without taking into account
the Pauli principle (QPM approach in following discussions).

\vspace{3mm}{\bf 3. Results and discussion}\vspace{1mm}

As an example we have performed numerical calculations for the $^{68}$Zn.
The Woods--Saxon potential parameters in use are from \cite{KVC96}, slightly
modified to better describe the ground state density. In so far as our
single-particle spectrum includes the bound and quasibound states we do not
use any effective charge to calculate the electromagnetic transition
probabilities. The pairing constants $G_\tau^{(0)}$ are fixed so as to
reproduce the odd-even mass difference of neighbouring nuclei. The strength
parameters $\kappa^{(\lambda)}$ for the cases based on the RPA and ERPA
schemes are adjusted so that the $B(E\lambda )$ values calculated
with the wave function (23) within both approaches are reasonably close
to the experimental ones. This means that $\kappa^{(\lambda)}$ for the ERPA
calculations are larger than for the RPA ones \cite{KVC94}. No changes of $%
\kappa^{(\lambda)}$ have been done for calculations without the two-phonon
terms. We use the ratio $\kappa_{1}^{(\lambda)}/\kappa_{0}^{(\lambda) }=-1.2$
that enables one to reproduce excitation energies of the isovector giant
resonances in spherical nuclei. It is worth to mention that in our previous
papers \cite{KVC93,KVC94,KVC96} we took into account only the isoscalar
interaction for the p-h channel, but the inclusion of the isovector
interaction does not affect the structure of low-lying states practically.

Our studies in Zn isotopes \cite{KVC94,KVC96} of the effect of coupling of
vibrations with different multipolarities ($\lambda $)(via $q_{jj^{\prime}}$
) show that in realistic calculations one can keep only $\lambda $=2 and 3
mainly. Solving nonlinear equations (9)-(14) one can find the phonon
amplitudes,energies and the quasiparticle distributions within ERPA. Making
use of them as input values it is possible to define from equations
(24),(31)-(33) energies and the structure of the states described by the
wave function (23).

Knowing the wave functions it is not difficult to calculate any matrix
elements and physical quantities. For example, the charge transition density
is calculated by the formula:
\begin{eqnarray}
\rho^{(J)}_{\nu}(r)=\sum_{j_{1}j_{2}}\rho^{J }_{j_{1}j_{2}}(r)\{\frac{1}{2}
(1\,-\,q_{j_{1}j_{2}})u^{(+)}_{j_{1}j_{2}}\, \sum_{i}R_{i}(J\nu)
(\psi^{Ji}_{j_{1}j_{2}}\,+\,\phi^{Ji}_{j_{1} j_{2}})\,-  \nonumber \\
v^{(-)}_{j_{1}j_{2}}\sum_{\lambda_{1} i_{1} \lambda_{2} i_{2}}\sqrt{
(2\lambda_{1}+1)(2\lambda_{2}+1)} P^{\lambda_{1}i_{1}}_{\lambda_{2}i_{2}}(J%
\nu) \sum_{j_{3}}(1\,-\,q_{j_{3}j_{1}})\times  \nonumber \\
\left\{
\begin{array}{ccc}
\lambda _1 & \lambda _2 & J \\
j_1 & j_2 & j_3
\end{array}
\right\} (\psi _{j_2j_3}^{\lambda _1i_1}\phi _{j_3j_1}^{\lambda _2i_2}+\psi
_{j_3j_1}^{\lambda _2i_2}\phi _{j_2j_3}^{\lambda _1i_1})\}  \label{eq34}
\end{eqnarray}
The expression for the two-quasiparticle transition density $%
\rho^{J}_{j_{1}j_{2}}(r)$ can be found in\cite{San91}. Our charge transition
densities are folded with the formfactor of the proton charge distribution
\cite{Sim80}. Using quantities (34) one calculates the reduced transition
probabilities from the ground to the excited state ($J\nu$)\cite{Sat83}
\begin{equation}
B(EJ ;0^{+}\rightarrow (J\nu))=(2J+1)\mid \int\limits_0^\infty
r^{J+2}\rho_{\nu}^{(J)}(r)dr\mid ^2.  \label{eq35}
\end{equation}

The results of our calculations for the quasiparticle distribution in the
ground state of $^{68}$Zn are shown in the table 1. It contains the values
of $q_j$ obtained in the RPA and ERPA schemes. As one can see from table 1
the $q_j$ have large values for the subshells near the Fermi surface only
and the ERPA gives stronger correlations in comparison with the RPA. A
similar behaviour of $q_j$'s has been found for other Zn isotopes \cite
{KVC94}. This is valid for our choice of the multipole constants. As it was
mentioned above the multipole constants $\kappa ^{(\lambda )}$ have been
chosen to describe with a reasonable accuracy experimental $B(E\lambda )$%
-values. The value of $\kappa^{(2)}_{0}=0.0259 $ MeV/fm for the non--linear
problem is quite larger than the critical RPA constant $%
\kappa^{(2)}_{0}=0.0242$ MeV/fm where the RPA solution becomes complex.
(In the RPA case $\kappa^{(2)}_{0}=0.0227$ MeV/fm).
The octupole constant in
use are equal to $\kappa^{(3)}_{0}=0.0235$ MeV/fm for the RPA and $%
\kappa^{(3)}_{0}=0.0250$ MeV/fm for the ERPA calculations respectively. The
last value is smaller than the critical RPA constant for the octupole
vibrations in contrast to the quadrupole ones. All the calculations based on
the RPA phonons (RPA, QPM, QPMPP) have been performed with the RPA
constants as well all the calculations based on the ERPA phonons
(ERPA, EQPM, EQPMPP) have been done with the ERPA set of constants.
It is worth to mention that as in the case of metallic clusters (\cite{CPSG96})
our ERPA calculation gives also weaker correlations compared to the
RPA ones, if the same set of multipole constants is used in both
approximations, but in this case neither energies nor transition
probabilities can be reproduced within the ERPA or its modifications.

To study the influence of the GSC on the quasiparticle-phonon coupling we
calculated the structure of the low-lying states in $^{68}$Zn with the wave
function (23). Experimental data \cite{DeV87,N77} and results of our
calculations for the $2^+_{1,2}$ and $3^-_{1}$ states within different
approaches are shown in the table 2.

One can see from table 2 that the RPA and ERPA overestimate the energies and
fail to reproduce the transition probabilities for the $2^+_{2}$ and $%
3^-_{1} $ states. Taking into account coupling of the one- and two-phonon
components improves essentially the description of all states under
consideration. Besides the transitions to the ground state, one can
reproduce the B(E2)-value for the E2-transition between the first and the
second $2^+$ states. The inclusion of the Pauli principle corrections in
two-phonon terms changes to worse the description of the second $2^+$
state mainly. The EQPMPP describes energies and transition probabilities
better than the QPMPP. It is worth to point out that the B(E2)-values for
the E2-transition between the first and the second $2^+$ states depend
essentially on the two-phonon components of the wave function of the $2^+_2$
state and the last ones can be affected by the three-phonon terms, which are
out of the present consideration. One can conclude that the most consistent
approach from theoretical point of view (EQPMPP) where the Pauli principle
is taken into account in both one- and two-phonon terms, gives a rather good
description of experimental data in general.

The transition probabilities are the integral characteristics of the
vibrational states and they are less sensitive to the details of the nuclear
wave functions than the differential ones. As it was pointed out in our
previous papers \cite{KVC93,KVC94,KVC96}, GSC affect essentially the charge
transition densities. Being the spatial overlap between the ground state
wave function and the excited state wave function, the charge transition
density provides a good test for nuclear models. The surface nature of the
low-lying collective states predicted by calculations performed within the
Hartree-Fock (HF) approach with effective forces \cite{BT75} and the finite
Fermi systems theory \cite{FKS79} has been demonstrated in the experiments
on inelastic electron scattering from magic nuclei \cite{He84}. Experimental
and theoretical (based on the random phase approximation (RPA)) studies of
the charge transition densities \cite{San91,Kim91} of the low-lying states
in some spherical nuclei are in reasonable agreement, but the theory gives
too large fluctuations of the transition densities in the interior region.
All theoretical calculations in RPA, as in HF, give the same behaviour in
the nuclear interior, which indicates a systematic problem of a more
fundamental nature (a detailed discussion can be found in refs. \cite
{HB83,FP87}).

To test the developed approach we calculated the charge transition densities
using eq.(34) in $^{68}$Zn within different approaches. The figures 1 and 2
show the transition charge densities ($\rho _\nu ^{(J)}(r)$) from the ground
to the first $2^{+}$ state in $^{68}$Zn. The experimental data \cite{N77}
are presented as a shadowed area. Fig.1 shows results of calculations based
on the RPA phonons while calculations based on the ERPA phonons are
presented in fig.2.

The RPA reproduces the behaviour of the charge transition densities
qualitatively, but it overestimates the interior part of the $\rho
_1^{(2)}(r)$. As one can see from fig.1, the inclusion of the two-phonon
terms (see the QPM case) reduces the bump in the interior part of $\rho
_1^{(2)}(r)$ by 17\%. This is due to the reduction of the contribution of
the first term in eq. (34) ($R_i<1$). In spite of the dominance of the first
term, the second one gives an additional reduction of $\rho$ too. The Pauli
principle correction for the two-phonon terms (QPMPP - case) change results
not more than by 3\%.

Taking into account GSC beyond the RPA results in a suppression of interior
oscillations by 9\% in comparison with the RPA case (see fig.2). Such a
depletion is related with the Pauli blocking effect for the proton
two-quasiparticle configuration $\{2p_{3/2},2p_{3/2}\}$, which is mainly
responsible for the interior bump in the charge transition densities in the
Zn isotopes. According to our RPA calculations the proton two-quasiparticle
configuration $\{2p_{3/2},2p_{3/2}\}$ gives a contribution about 40\% into
the norm of the first quadrupole phonon in $^{68}$Zn. The inclusion of the
GSC redistributes the strength of this configuration over many phonon roots
and as a result the contribution into the first root becomes lower. The GSC
suppresses the contribution of the partial proton two-quasiparticle $%
\{2p_{3/2},2p_{3/2}\}$ transition density for which $q_{2p_{3/2}}$ has the
biggest value and as it was mentioned above plays an essential role in the
structure of the interior part of the transition density for the $2_1^{+}$
states. The configuration $\{1f_{5/2},1f_{5/2}\}$ gives some contribution to
the interior part too and the same mechanism of suppression takes place for
it. It should be noted that the amplitudes of the oscillations for the
configurations with low orbital momenta are bigger than for the ones with
high orbital momenta. That is because the single particle wave functions
with low orbital momenta are mainly localized in the interior part of
nuclei. Besides the blocking effect there are changes in coefficients $%
u_{j_1j{2}}^{(+)}$ because of the influence of the GSC on pairing and in the
phonon amplitudes. All these effects suppress the interior bump. Taking into
account the phonon coupling we get an additional lowering of the interior
bump in the charge transition density and a reason of such lowering is the
same as it was discussed above for the coupling of the RPA phonons. Finally
the EQPM approach gives 30\% reduction of the $\rho _1^{(2)}(r)$ in the
interior nucleus region. It is worth to note that the Pauli principle
corrections in two-phonon terms change the results slightly, but such
corrections must be taken into account because they are often responsible
for the weak electromagnetic transitions between excited states. One can see
from eq.(35) that the outer part of the charge transition density is
responsible mainly for the value of the reduced transition probability.
Since all above discussed approaches have very close values for this part of
the charge transition density, the calculated B(E2) are close too.

We would like to emphasize that the depletion effect discussed above can not
be reproduced by any renormalization of the force strength $\kappa ^{(2)}$
in the RPA if one wants to describe energies and reduced transition
probabilities. An influence of the p-p channel of the multipole forces on
the charge transition densities is under an investigation now. As concerning
the charge transition density for the second $2^+$ state, it is impossible
to treat it without taking into account the three-phonon terms, which are
out of the present consideration due to its numerical complexity.

The charge transition densities for the $3_1^{-}$ states in all Zn isotopes
have a clear surface nature and there are no strong oscillations in the
interior region of the nucleus because of a destructive interference of the
2-qp partial transition densities constructed from the single-particle wave
functions with different parity.

\vspace{3mm} {\bf 4.Conclusion}\vspace{1mm}

A consistent treatment of the ground state correlations beyond the RPA
including their influence on the pairing and phonon-phonon coupling is
presented. A new general system nonlinear equations for the quasiparticle
phonon model is derived. It is shown that this system contains as a
particular case all equations derived for the QPM early. The system is
solved numerically for first time in a realistic case for $^{68}$Zn to study
the effect of the GSC on the excitation energies, transition probabilities
and charge transition densities of the vibrational states. Taking into
account the GSC results in better agreement with experimental data for the
characteristics of the low-lying states.

Up to now all realistic calculations taking into account GSC beyond the RPA
have been performed with separable effective forces only (see \cite
{KVC93,KVC94,TS97,HHC97}) because even for such simplified forces it is not
very easy to solve the set of nonlinear equations in a large configuration
space. The finite rank separable approximation for the Skyrme forces
suggested recently \cite{GSV98} opens a new possibility to treat GSC beyond
RPA basing on more realistic nuclear interactions.

\vspace{3mm} {\bf 5. Aknoweledgments}

This work has been partially supported by the Bulgarian National Science
Foundation under contract $\Phi $--621 and the Russian Foundation for Basic
Researches under grant 96-15-96729. Two of us (D.K. and V.V.V.) thank the
INFN for the hospitality and support during their stay in Catania, where a
part of this work was done.

\newpage

\begin{table}[th]
\caption{ Quasiparticle distribution in the ground state of $^{68}$Zn}
\label{tab1}
\begin{center}
\begin{tabular}{|c|cc|cc|}
\hline
& \multicolumn{4}{c|}{$q_{nlj}$} \\[1mm] \hline
& \multicolumn{2}{c|}{neutrons} & \multicolumn{2}{c|}{protons} \\[1mm]
nlj & {\scriptsize RPA} & {\scriptsize ERPA} & {\scriptsize RPA} &
{\scriptsize ERPA} \\[1mm] \hline
$2s_{1/2}$ & 0.0047 & 0.0134 & 0.0122 & 0.0189 \\[1mm]
$1f_{7/2}$ & 0.0163 & 0.0222 & 0.0320 & 0.0451 \\[1mm]
$2p_{3/2}$ & 0.0540 & 0.0651 & 0.0862 & 0.1053 \\[1mm]
$1f_{5/2}$ & 0.0482 & 0.0609 & 0.0324 & 0.0445 \\[1mm]
$2p_{1/2}$ & 0.0899 & 0.0994 & 0.0359 & 0.0461 \\[1mm]
$1g_{9/2}$ & 0.0230 & 0.0295 & 0.0144 & 0.0161 \\[1mm]
$2d_{5/2}$ & 0.0072 & 0.0087 & 0.0042 & 0.0047 \\[1mm] \hline
\end{tabular}
\end{center}
\end{table}

\begin{table}[th]
\caption{Energies and B(E$\lambda $)-values for up-transitions to some first
vibrational states of $^{68}$Zn}
\label{tab2}
\begin{center}
\begin{tabular}{|c|c|c|c|c|c|c|}
\hline
State & \multicolumn{2}{|c}{2$_1^{+}$} & \multicolumn{2}{|c}{2$_2^{+}$} &
\multicolumn{2}{|c|}{3$_1^{-}$} \\[1mm] \hline
& Energy & B(E2) & Energy & B(E2) & Energy & B(E3) \\[1mm] \hline
&  & 0$^{+}\rightarrow 2_1^{+}$ &  & 0$^{+}\rightarrow 2_2^{+}$ &  & 0$%
^{+}\rightarrow 3_1^{-}$ \\[1mm]
&  &  &  & 2$_1^{+}\rightarrow 2_2^{+}$ &  &  \\[1mm] \hline
& (MeV) & (e$^2$fm$^4$) & (MeV) & (e$^2$fm$^4$) & (MeV) & (e$^2$fm$^6$) \\
EXP. & 1.077 & 1266 & 1.883 & 46 & 2.751 & 38400 \\[1mm]
&  &  &  & 287 &  &  \\[1mm] \hline
RPA & 1.360 & 1290 & 2.390 & 75 & 3.830 & 48240 \\[1mm]
&  &  &  &  &  &  \\[1mm] \hline
QPM & 1.090 & 1200 & 2.060 & 48 & 2.760 & 38550 \\[1mm]
&  &  &  & 231 &  &  \\[1mm] \hline
QPMPP & 1.140 & 1220 & 2.180 & 27 & 2.840 & 39200 \\[1mm]
&  &  &  & 351 &  &  \\[1mm] \hline
ERPA & 1.330 & 1250 & 2.320 & 106 & 3.980 & 43070 \\[1mm]
&  &  &  &  &  &  \\[1mm] \hline
EQPM & 1.080 & 1170 & 1.810 & 47 & 2.760 & 34700 \\[1mm]
&  &  &  & 212. &  &  \\[1mm] \hline
EQPMPP & 1.080 & 1270 & 1.960 & 38 & 2.750 & 35660 \\[1mm]
&  &  &  & 436 &  &  \\[1mm] \hline
\end{tabular}
\end{center}
\end{table}

\begin{figure}[p]


\centerline{\epsffile{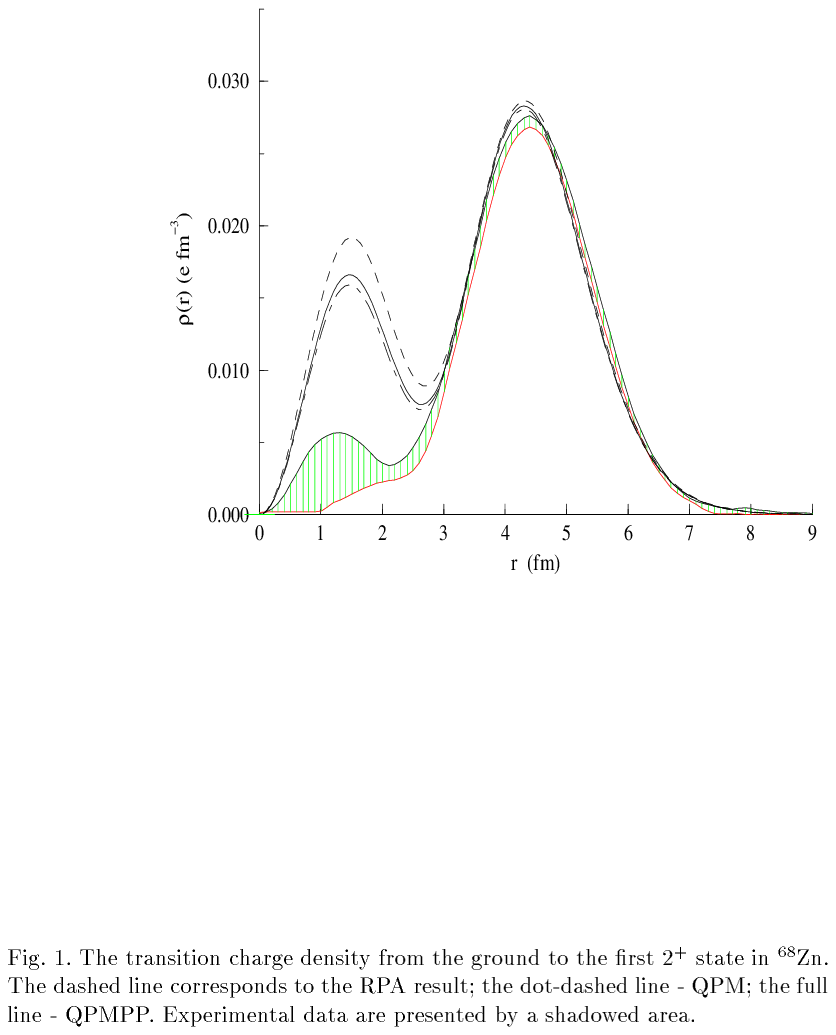}}

\end{figure}

\begin{figure}

\centerline{\epsffile{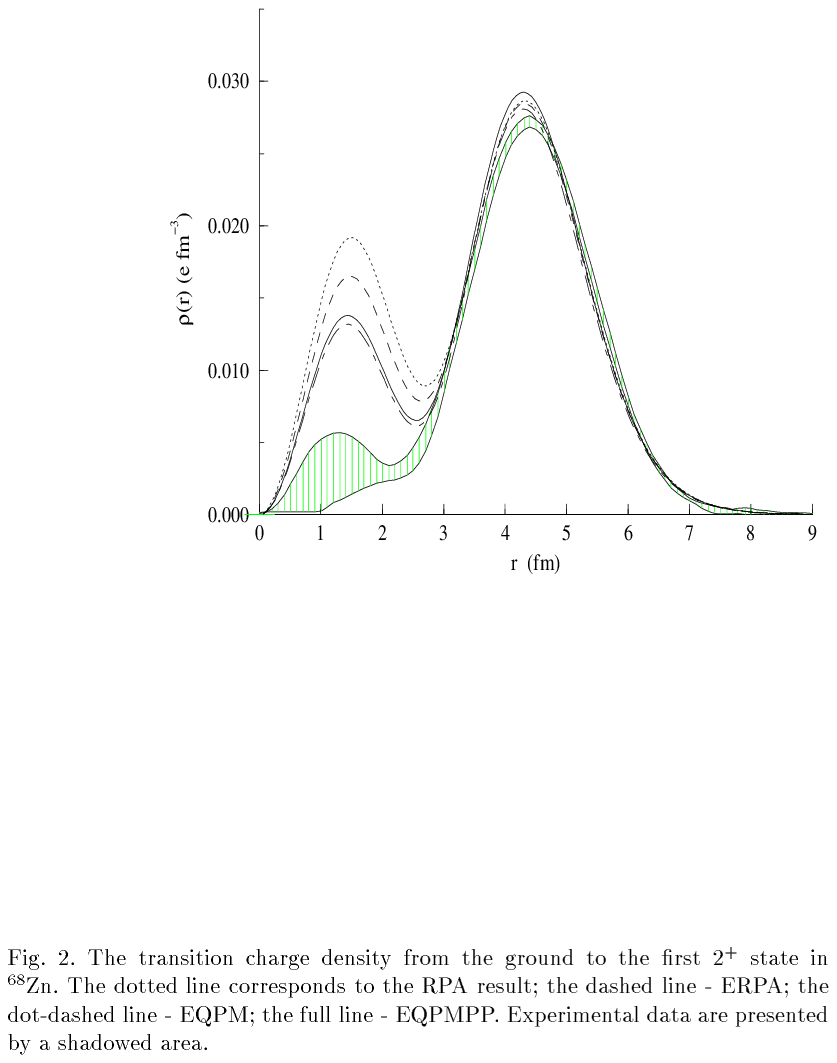}}
\end{figure}

\end{document}